\documentclass[a4paper, 10pt,conference]{IEEEtran}
\IEEEoverridecommandlockouts

\usepackage{placeins}
\usepackage{cite}
\usepackage{amsmath,amssymb,amsfonts}
\usepackage{newtxmath}
\usepackage{algorithmic}
\usepackage{graphicx}
\usepackage{booktabs}
\usepackage{textcomp}
\usepackage{xcolor}
\definecolor{myorange}{RGB}{255, 166, 106}
\definecolor{mygreen}{RGB}{34, 177, 76}
\usepackage{soul}
\def\BibTeX{{\rm B\kern-.05em{\sc i\kern-.025em b}\kern-.08em
    T\kern-.1667em\lower.7ex\hbox{E}\kern-.125emX}}

\usepackage[hyphens]{url}
\usepackage{breqn}

\usepackage{caption}
\usepackage{subcaption}

\usepackage{glossaries}
\usepackage{float}

\makeglossaries

\newacronym{mmwave}{mmWave}{millimeter-wave}
\newacronym{siso}{SISO}{single-input single-output}
\newacronym{miso}{MISO}{multiple-input single-output}
\newacronym{simo}{SIMO}{single-input multiple-output}
\newacronym{mimo}{MIMO}{multiple-input multiple-output}
\newacronym{tx}{Tx}{transmitter}
\newacronym{rx}{Rx}{receiver}
\newacronym{rf}{RF}{radio frequency}
\newacronym{mpc}{MPC}{multipath component}
\newacronym{ula}{ULA}{uniform linear array}
\newacronym{aoa}{AOA}{angle-of-arrival}
\newacronym{aod}{AOD}{angle-of-departure}
\newacronym{cir}{CIR}{channel impulse response}
\newacronym{pa}{PA}{power amplifier}
\newacronym{if}{IF}{intermediate frequency}
\newacronym{lo}{LO}{local oscillator}
\newacronym{clk}{CLK}{clock reference}
\newacronym{dac}{D/A}{digital-to-analog converter}
\newacronym{zc}{ZC}{Zadoff-Chu}
\newacronym{fpga}{FPGA}{Field-Programmable Gate Array}
\newacronym{los}{LOS}{Line-Of-Sight}
\newacronym{nlos}{NLOS}{Non Line-Of-Sight}
\newacronym{hpbw}{HPBW}{half-power beam width}
\newacronym{subthz}{sub-THz}{sub-terahertz}
\newacronym{vaa}{VAA}{virtual antenna array}
\newacronym{dss}{DSS}{directional scanning scheme}
\newacronym{snr}{SNR}{signal-to-noise-ratio}
\newacronym{uca}{UCA}{uniform circular array}
\newacronym{ura}{URA}{uniform rectangular array}
\newacronym{sage}{SAGE}{space-alternating generalized expectation-maximization}
\newacronym{vna}{VNA}{vector network analyzer}
\newacronym{lna}{LNA}{low-noise amplifier}
\newacronym{pdp}{PDP}{power-delay profile}
\newacronym{pmf}{PMF}{polymer microwave fiber}
\newacronym{ap}{AP}{access point}
\newacronym{ru}{RU}{radio unit}
\newacronym{hrpe}{HRPE}{high-resolution parameter estimation}
\newacronym{isac}{ISAC}{integrated sensing and communication}
\newacronym{rms}{RMS}{root mean square}

\begin{document}
\bstctlcite{BSTcontrol}

\title{On the Feasibility of Human Presence Detection
Using Ceiling-Mounted Sub-THz Channel
Sounding: Conference Room Measurement

\thanks{This work has been funded by the Swedish Research Council (Grant No. 2022-04691), the Horizon Europe Framework Programme under the Marie Sk{\l}odowska-Curie grant agreement No. 101059091, the Horizon 2020 EU Framework Programme under Grant Agreement No.\,861222, the Royal Physiographic Society of Lund, the Strategic Research Area Excellence Center at Link\"oping--Lund in Information Technology (ELLIIT), and Ericsson.
}
}

\author{
\IEEEauthorblockN{Ali Al-Ameri*, Xuesong Cai‡, Aleksei Fedorov*, and Fredrik Tufvesson*}
\IEEEauthorblockA{*Department of Electrical and Information Technology, \textit{Lund University}, Lund, Sweden \\ ‡State Key Laboratory of Photonics and Communications, School of Electronics, \textit{Peking University}, Beijing, China \\
 \{ali.al-ameri, aleksei.fedorov@eit.lth.se, fredrik.tufvesson\}@eit.lth.se \\ xuesong.cai@pku.edu.cn
\footnotesize \textsuperscript{}
}
}

\maketitle

\begin{abstract}
This paper presents a measurement-based investigation on the feasibility of human presence detection using a ceiling-mounted sub-THz channel sounder operating from 134 to 146~GHz. Wideband channel measurements were conducted in an indoor conference room under empty-room, human-present, and water-filled mannequin scenarios across five spatial positions. The measurements were performed using a vector network analyzer combined with sub-THz frequency extenders. Two antenna beamwidth configurations were used, one with a highly directive horn antenna on the transmitter side and one with a less directive, open-waveguide transmitter. The measured channel responses were transformed into calibrated power delay profiles and analyzed using normalized channel variation metrics in the delay domain. The results show that human detection is strongly dependent on target position relative to the ceiling-mounted transmitter and receiver as well as on antenna beamwidth. Furthermore, repeated empty-room measurements reveal that small environmental changes, such as slight furniture displacement, introduce non-negligible channel variations that must be considered when evaluating detection performance. In the wide-beam open-waveguide configuration, the human-present measurements produced lower values of the delay-domain variation metric than the repeated empty-room baseline, whereas the water-filled mannequin produced values at or above this baseline across all positions. With the directive transmitter, the human response exceeded the baseline significantly but only at favorable positions, especially P1 and P2, showing that the sensing response remains spatially selective. These findings provide experimental insight into the capabilities and limitations of ceiling-mounted sub-THz sensing for future integrated sensing and communication systems.
\vspace{0.3cm}
\end{abstract}

\begin{IEEEkeywords}
Sub-THz sensing, human presence detection, ISAC, ceiling-mounted channel sounding, and wideband channel measurements.
\end{IEEEkeywords}

\section{Introduction}

The sub-terahertz (sub-THz) frequency range is expected to play an important role in future wireless communication systems because of the large available bandwidths and narrow beams \cite{8732419,9766110,10439212}. These characteristics also make sub-THz systems attractive for \gls{isac}, where the same radio infrastructure is used for both communication and sensing of the surrounding environment.

One important sensing task in indoor \gls{isac} systems is human presence detection \cite{7765354}. Such functionality can support energy-efficient building control, occupancy awareness, and safety monitoring \cite{9569364}. In this context, channel-based sensing is particularly attractive because it infers the presence of a target from changes in the communication channel itself, rather than relying on dedicated radar hardware. This is especially relevant at sub-THz frequencies, where the hardware is expensive and it is therefore desirable to reuse the same infrastructure for both communication and sensing \cite{10707308}.

At sub-THz frequencies, the human body interacts strongly with electromagnetic waves because of its high water content and lossy dielectric properties \cite{8732419,9569364,6574880}. A human subject can attenuate existing propagation paths, introduce additional scattering, and redistribute energy in the delay domain. However, the same sensitivity that makes sub-THz sensing promising also makes it challenging. Small environmental changes, such as slight displacement of furniture, movement of cables, or changes in surrounding objects, may also introduce significant disturbances in the measured channel. Reliable human detection therefore requires distinguishing target-induced changes in the measured channel from background channel variations.

A practically relevant scenario for indoor deployment is a ceiling-mounted sensing system. Ceiling-mounted sensing systems, and more generally ceiling-mounted radio infrastructure, are attractive because they can provide wide-area coverage without occupying floor space while also reducing blockage from furniture and other obstacles \cite{11225925,8986750,8739348}. However, the propagation characteristics in such vertical links differ from those of conventional horizontal links. In a downward-looking configuration, the received signal is influenced not only by \gls{los} propagation, but also by floor reflections, interactions with tables and surrounding objects, diffuse multipath components, and human-body-induced channel variations \cite{8739348,9790802}. As a result, target detection cannot be characterized solely as a simple \gls{los} blockage problem. It is therefore important to investigate experimentally how target position and antenna beamwidth influence detection performance and channel behavior in such ceiling-mounted geometries for sub-THz frequencies \cite{8986750}.

To this end, this paper presents a measurement-based analysis of human presence detection using a ceiling-mounted sub-THz channel sounder operating over a 12~GHz bandwidth from 134 to 146~GHz. The measurements were performed in a downward-looking bistatic configuration, where both the \gls{tx} and \gls{rx} were mounted on the ceiling and pointed toward the ground. The measurement campaign included repeated empty-room measurements, human-present measurements, and water-filled mannequin measurements at five different positions in the room. Two antenna configurations were considered: an open-waveguide configuration and a directive-transmitter configuration. The measured channel responses were transformed into calibrated \glspl{pdp} and analyzed through delay-domain variation metrics relative to repeated empty-room references.

The remainder of the paper is organized as follows. Section~II describes the measurement setup, environment, and scenarios. Section~III presents the processing procedure and the delay-domain metrics used for analysis. Section~IV discusses the measurement results, and Section~V concludes the paper.

\section{Measurement Setup, Environment and Scenarios}

The measurement campaign was carried out in a furnished indoor conference room. The channel sounder consisted of a \gls{vna} connected to sub-THz frequency extenders in a similar fashion to \cite{10152010,9124887}. The system measured the complex transfer function $S_{21}(f)$ over the frequency interval from 134 to 146~GHz. The bandwidth was therefore $B=12$~GHz and the number of frequency samples was $N_f=1001$, resulting in a delay resolution of approximately $\Delta d = c/B \approx 2.5$~cm. The corresponding maximum observable equivalent propagation distance was approximately $d_{\max} = c/\Delta f \approx 25$~m, where $c$ is the speed of light, and $\Delta f = B/(N_f-1)$ is the frequency spacing. The \gls{if} bandwidth was set to 10~Hz in order to obtain a high measurement dynamic range.

\begin{figure}[!h]
    \centering
    
    \begin{subfigure}[b]{0.95\linewidth}
        \centering
        \includegraphics[width=\linewidth]{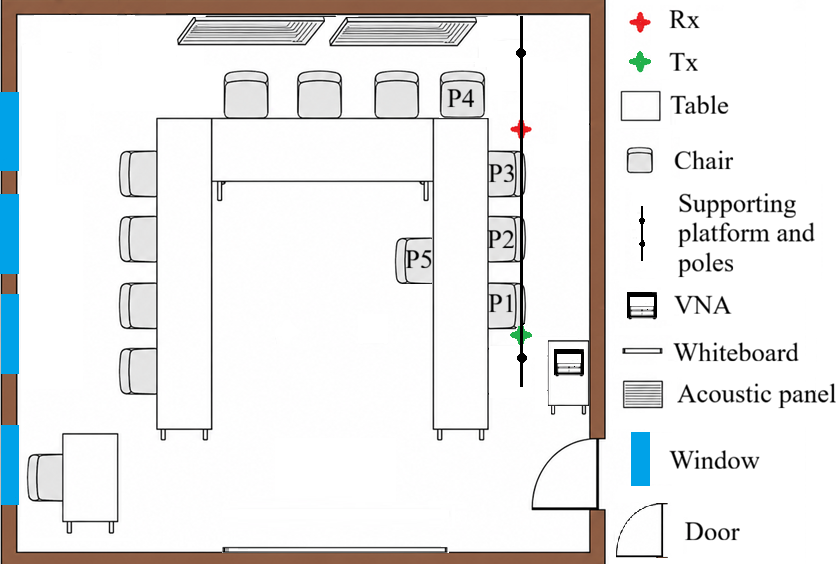}
        \caption{}
        \label{fig:2dsetup}
    \end{subfigure}
    
    \vspace{0.5cm}
    
    \begin{subfigure}[b]{0.95\linewidth}
        \centering
        \includegraphics[width=\linewidth]{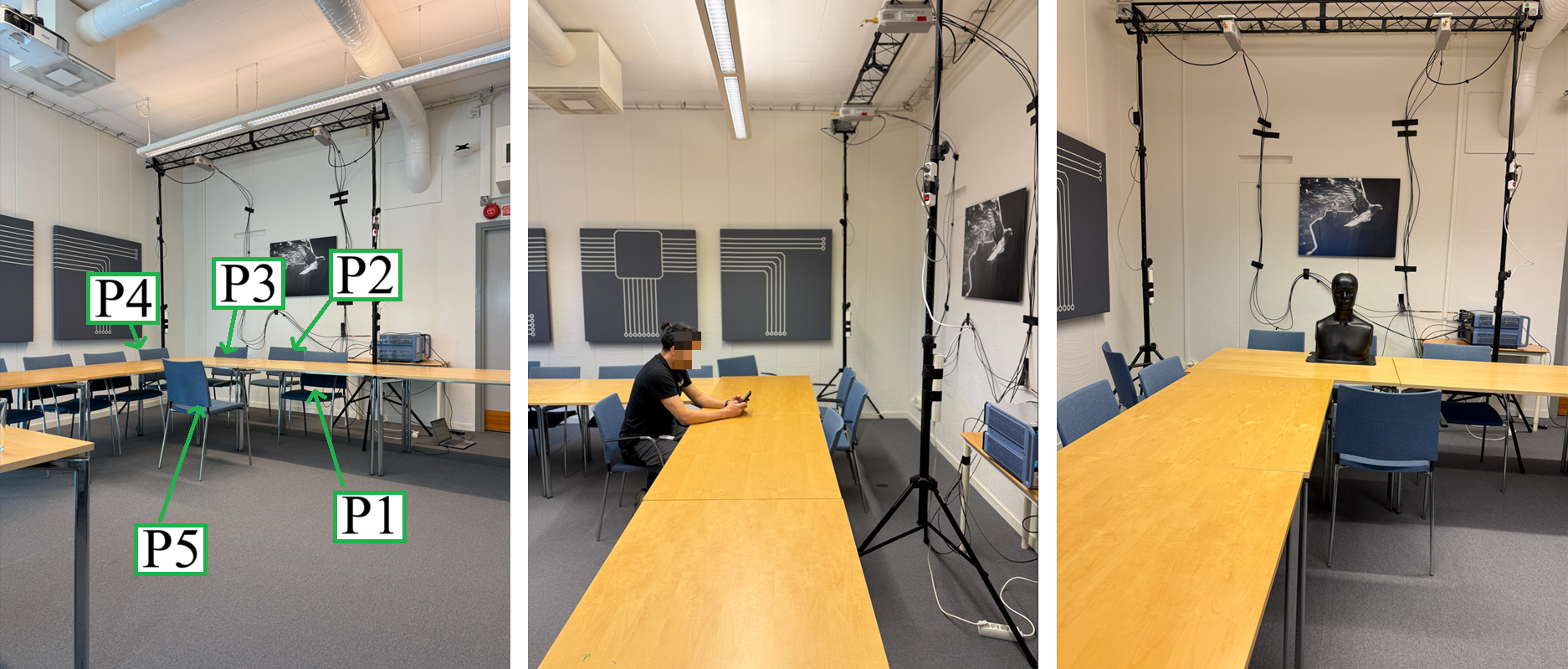}
        \caption{}
        \label{fig:irlPIC}
    \end{subfigure}
    
   \caption{(a) A 2D sketch of the ceiling-mounted sub-THz measurement setup and target positions P1--P5. The same position indices were used for both the human and water-filled mannequin measurements. (b) Photographs of the measurement environment. Left: Empty-room scenario. Middle: Human-present scenario with the subject located at position P5. Right: Water-filled mannequin scenario with the mannequin located at position P2.}
    \label{fig:measurementSetupAndEnvironment}
\end{figure}

The conference room in which the measurements took place contained tables, chairs, whiteboards, windows, and acoustic panels, as seen in Fig.~\ref{fig:2dsetup} and Fig.~\ref{fig:irlPIC}. Both the transmitter and receiver were mounted on the ceiling using a metallic platform supported by metallic poles. The frequency extender heads were oriented parallel to the floor, while 90$^\circ$ waveguide bends were used to direct the transmit and receive antennas downward, pointing toward the floor, forming a downward-looking bistatic sensing geometry. The measured response therefore includes contributions from floor and table reflections, scattering from surrounding furniture, and channel variations caused by the human body or mannequin. 

Two antenna configurations were evaluated. In the first configuration, open-waveguides were used at both the transmitter and receiver, resulting in relatively broad room illumination due to their approximate \gls{hpbw} of 62$^\circ$/94$^\circ$ in the E/H planes. In the second configuration, the transmitter was equipped with a directive horn antenna with an approximate beamwidth of 25$^\circ$/29$^\circ$ in the E/H planes, while the receiver again used an open-waveguide. This enables us to investigate the impact of different antenna beamwidths on detection performance of ceiling-mounted sensing systems.

Measurements were conducted at five target positions, denoted P1--P5 in Fig.~\ref{fig:measurementSetupAndEnvironment}, corresponding to different seating locations around the conference table. These positions represent different spatial relationships with respect to the ceiling-mounted transmitter and receiver, with some located within the main illuminated region of the antennas and others more peripherally. This spatial diversity is important for assessing how strongly the detection performance depends on the target location relative to the \gls{tx} and \gls{rx}. 

The measurements were first carried out for the open-waveguide configuration. In this case, the measurement sequence was as follows: an empty-room measurement, followed by human-present measurements at P1--P5, a second empty-room measurement, and then water-filled mannequin measurements at P1--P5. After completing these measurements, the transmitter was replaced with the directive antenna, and a second measurement sequence was performed as follows: an empty-room measurement, followed by human-present measurements at P1--P5, and a final empty-room measurement. Thus, mannequin measurements were only conducted for the open-waveguide configuration. During each recording, the subject or mannequin remained stationary while the channel response was measured. This repeated-reference structure is important for the processing in Section~III and for the interpretation of the detection metrics in Section~IV.

Before describing the measurement data processing procedure in the next section, the main parameters of the measurement campaign are summarized in Table~\ref{tab:measurement_parameters}.

\begin{table}[!t]
\caption{Measurement Parameters}
\centering
\setlength{\tabcolsep}{3pt}
\begin{tabular}{@{} l l @{}}
\hline
Parameter & Value \\
\hline
Frequency range & 134--146 GHz \\
Center frequency & 140 GHz \\
Bandwidth & 12 GHz \\
Frequency points & 1001 \\
IF bandwidth & 10 Hz \\
Geometry & Ceiling-mounted bistatic \\
Antenna configuration 1 & TX/RX open WG, HPBW $\approx 62^\circ/94^\circ$ (E/H) \\
Antenna configuration 2 & TX horn, RX open WG, HPBW $\approx 25^\circ/29^\circ$ (E/H) \\
Positions & 5 (P1--P5) \\
Scenarios & Empty room, human present, mannequin present \\
\hline
\end{tabular}
\label{tab:measurement_parameters}
\end{table}

\section{Processing}

For each measurement, the complex transfer function $S_{21}(f)$ over the band of interest is recorded. To reduce the influence of the measurement hardware, a back-to-back calibration was first performed for each antenna configuration before the over-the-air measurements. The calibrated channel response is obtained as
\begin{equation}
\tilde{S}_{21}(f_k)=\frac{S_{21}^{\mathrm{meas}}(f_k)}{S_{21}^{\mathrm{b2b}}(f_k)},
\label{eq:calibration}
\end{equation}
where $S_{21}^{\mathrm{meas}}(f_k)$ is the measured response and $S_{21}^{\mathrm{b2b}}(f_k)$ is the corresponding back-to-back reference. After this step, the remaining variations are primarily attributed to the propagation channel rather than to the frequency response of the measurement setup.

To obtain a delay-domain representation, a Hann window was applied to the calibrated frequency response before transformation,
\begin{equation}
\tilde{S}_{21}^{(w)}(f_k)=\tilde{S}_{21}(f_k)w_k,
\label{eq:windowedS21}
\end{equation}
where $w_k$ denotes the window coefficients. This reduces sidelobes in the delay domain, but also means that the measured delay response is the true physical channel convolved with the delay-domain response of the finite measurement bandwidth and window.

The windowed delay-domain response is obtained as
\begin{equation}
\hat{h}(\tau_n)=\mathrm{IFFT}\{\tilde{S}_{21}^{(w)}(f_k)\}
=\frac{1}{N_f}\sum_{k=0}^{N_f-1}\tilde{S}_{21}^{(w)}(f_k)e^{j2\pi f_k\tau_n},
\label{eq:ifft}
\end{equation}
and the corresponding measured \gls{pdp} is defined as
\begin{equation}
P(\tau_n)=|\hat{h}(\tau_n)|^2.
\label{eq:pdp}
\end{equation}

The underlying physical channel can be written conceptually as a superposition of multipath components,
\begin{equation}
h(\tau)=\sum_{\ell=1}^{L}\beta_\ell \delta(\tau-\tau_\ell),
\label{eq:channel_model}
\end{equation}
where $\beta_\ell$ is the complex path coefficient, including the phase term associated with the carrier frequency, and $\tau_\ell$ is the delay of the $\ell$th component. Due to the finite measurement bandwidth and the applied Hann window, the measured delay response is instead
\begin{equation}
\hat{h}(\tau)=h(\tau)*g_w(\tau)
=\sum_{\ell=1}^{L}\beta_\ell g_w(\tau-\tau_\ell),
\label{eq:windowed_channel_model}
\end{equation}
where $g_w(\tau)$ is the delay-domain spreading function associated with the measurement bandwidth and window. In the considered ceiling-mounted scenario, the presence of a human or mannequin may attenuate existing paths, introduce additional scattered components, or redistribute energy among delayed multipath contributions. In terms of the measured windowed delay response, a target-present measurement can therefore be written conceptually as
\begin{equation}
\hat{h}_{\mathrm{tar}}(\tau)=\hat{h}_{\mathrm{emp}}(\tau)
+\Delta \hat{h}_{\mathrm{tar}}(\tau)+\eta(\tau),
\label{eq:target_model}
\end{equation}
where $\hat{h}_{\mathrm{emp}}(\tau)$ is the measured empty-room response, $\Delta \hat{h}_{\mathrm{tar}}(\tau)$ is the target-induced change, and $\eta(\tau)$ represents residual measurement noise and environmental drift.

The repeated empty-room measurements described in Section~II showed that the background channel was not perfectly static. For this reason, the analysis is based on differences relative to empty-room references rather than on absolute target-present responses alone. Following the \gls{pdp} definition in \eqref{eq:pdp} and the target-present model in \eqref{eq:target_model}, let $P_{\mathrm{tar}}(\tau_n)$ denote the \gls{pdp} of a human-present or mannequin-present measurement, and let $P_{\mathrm{emp}}(\tau_n)$ denote the corresponding empty-room \gls{pdp}. Based on this, one useful metric is the normalized \gls{pdp} difference
\begin{equation}
D=\frac{\sum_n |P_{\mathrm{tar}}(\tau_n)-P_{\mathrm{emp}}(\tau_n)|}
{\sum_n P_{\mathrm{emp}}(\tau_n)}.
\label{eq:D_metric}
\end{equation}
This metric was chosen because it captures changes in both the shape of the \gls{pdp} and the delay-wise distribution of received power, rather than only changes in total power. This is particularly relevant here, since human presence may alter several multipath contributions without necessarily causing a large overall attenuation. Unlike correlation-based measures, which primarily quantify similarity in overall \gls{pdp} shape, the metric in \eqref{eq:D_metric} directly measures the magnitude of the delay-domain change relative to the empty-room case. It is therefore better aligned with the considered detection problem, where target-induced responses must be assessed against the baseline variation observed between repeated empty-room measurements.

This metric in \eqref{eq:D_metric} was applied between repeated empty-room measurements in order to quantify the baseline channel variation of the environment. These empty-to-empty comparisons provide the reference level for interpreting target-present measurements in Section~IV. Importantly, target-induced behavior may appear either as an increase or as a systematic reduction of $D$ relative to this baseline, depending on how the target changes the delay-domain response compared with the background drift.

In addition to $D$, the total delay-domain energy was computed as
\begin{equation}
P_{\mathrm{tot}}=\sum_n P(\tau_n),
\label{eq:ptot}
\end{equation}
and the relative power change between target-present and reference measurements was defined as
\begin{equation}
\Delta P_{\mathrm{dB}}=10\log_{10}\frac{\sum_n P_{\mathrm{tar}}(\tau_n)}
{\sum_n P_{\mathrm{emp}}(\tau_n)}.
\label{eq:deltaP_metric}
\end{equation}
To further characterize how the multipath structure changes, the mean delay and \gls{rms} delay spread were computed as
\begin{equation}
\bar{\tau}=\frac{\sum_n \tau_n P(\tau_n)}{\sum_n P(\tau_n)}
\label{eq:mean_delay}
\end{equation}
and
\begin{equation}
\sigma_\tau=
\sqrt{\frac{\sum_n(\tau_n-\bar{\tau})^2P(\tau_n)}
{\sum_nP(\tau_n)}}.
\label{eq:rms_delay}
\end{equation}
Together, these quantities enable target-induced changes to be assessed relative to empty-room variation and help distinguish between overall power attenuation and structural redistribution of multipath energy in the delay domain.

\FloatBarrier
\section{Results and Discussion}

The results are interpreted primarily through the normalized channel variation metric $D$ defined in \eqref{eq:D_metric}, while repeated empty-to-empty comparisons provide the empty-room baseline mentioned in Section~III. The first observation is that the background channel is not perfectly stable over the course of the measurement campaign. Repeated empty-room measurements yield baseline values of approximately $D=0.455$ for the open-waveguide configuration and $D=0.448$ for the directive-\gls{tx} configuration. These values quantify the channel variation caused by measurement-to-measurement changes in the nominally empty room. Consequently, target-present measurements should not be interpreted only by whether $D$ is large in an absolute sense, but by how their values deviate from this empty-room baseline and whether this deviation is systematic across positions or object types.

\begin{figure}[!h]
\centering
\includegraphics[width=0.95\columnwidth]{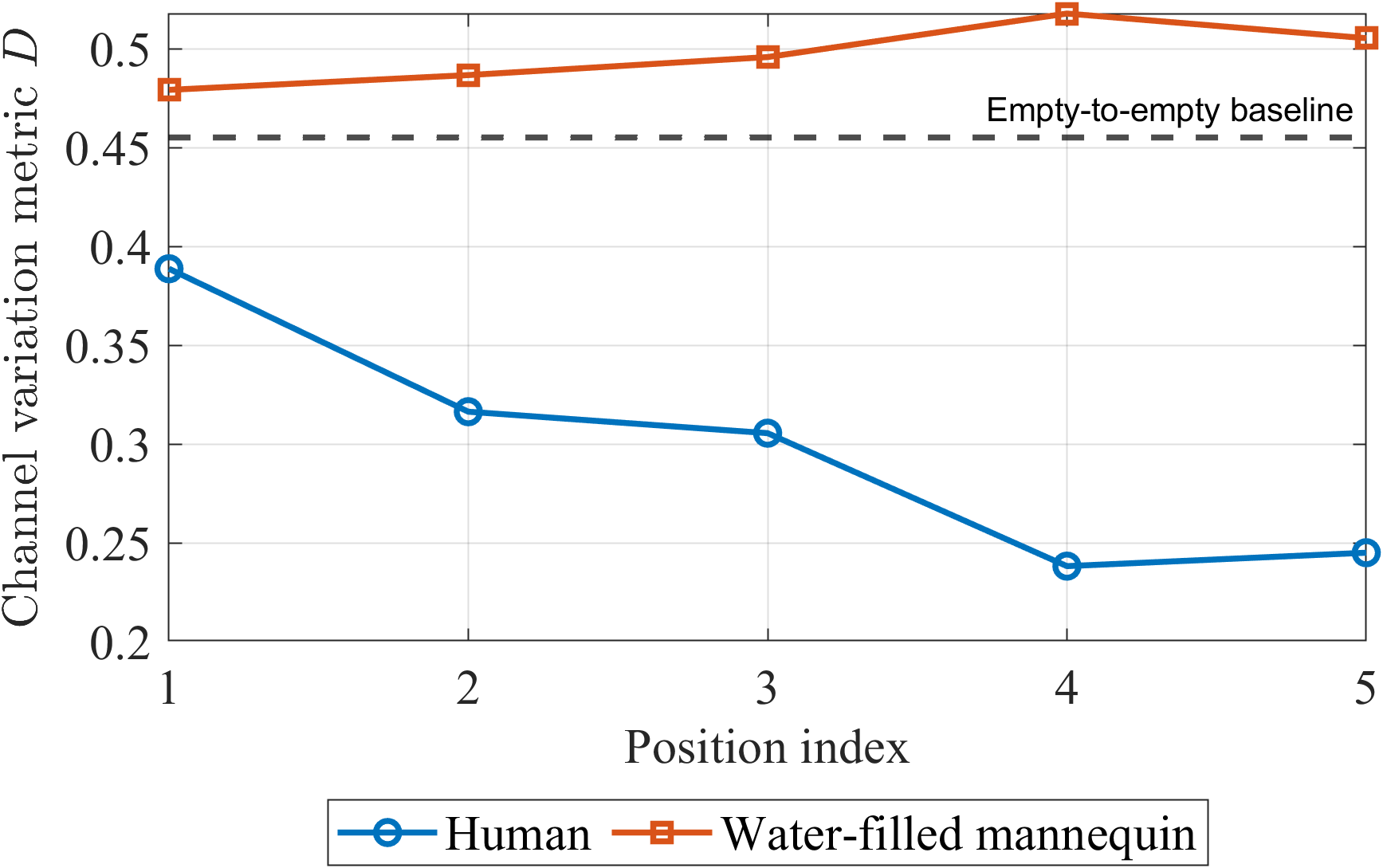}
\caption{Position-wise normalized channel variation metric $D$ for the open-waveguide configuration. The empty-to-empty baseline $D=0.455$ is shown for reference.}
\label{fig:openwg_metric}
\end{figure}

Fig.~\ref{fig:openwg_metric} shows the position-dependent behavior for the open-waveguide configuration, where the target positions P1--P5 are illustrated in Fig.~\ref{fig:measurementSetupAndEnvironment}. For the human-present case, the normalized channel variation metric decreases from approximately $D=0.389$ at P1 to about $D=0.24$--$0.25$ at P4 and P5. These values are below the empty-to-empty baseline of $D=0.455$, but this does not mean that the human-present measurements contain no sensing information. Rather, it indicates that the human-present PDPs are more similar to the chosen empty-room reference than the two repeated empty-room measurements are to each other. This systematic reduction of $D$ is itself informative and suggests that the human changes the channel in a different way from the background drift captured by the empty-to-empty comparison. However, because the effect appears as a reduction relative to the baseline rather than as a simple increase, one can conclude that delay-domain thresholding on $D$ alone is not sufficient for robust human detection.

The water-filled mannequin produces the opposite trend. Its values of the normalized channel variation metric remain between approximately $D=0.479$ and $D=0.518$ across all five positions, that is, consistently at or above the empty-to-empty baseline. This indicates a stronger interaction with the channel that increases the delay-domain difference relative to the empty-room reference. The opposite directions of the human and mannequin trends are therefore important: the human-present case tends to reduce $D$ relative to the empty-to-empty baseline, whereas the mannequin increases it. This suggests that the metric contains information about the type of object in the room, not only about the presence of any object. Nevertheless, the interpretation depends on the reference measurements and the measurement sequence, so again, a more reliable detector would need additional sensing domains rather than a single fixed threshold on $D$.

\begin{figure}[!h]
\centering
\includegraphics[width=0.95\columnwidth]{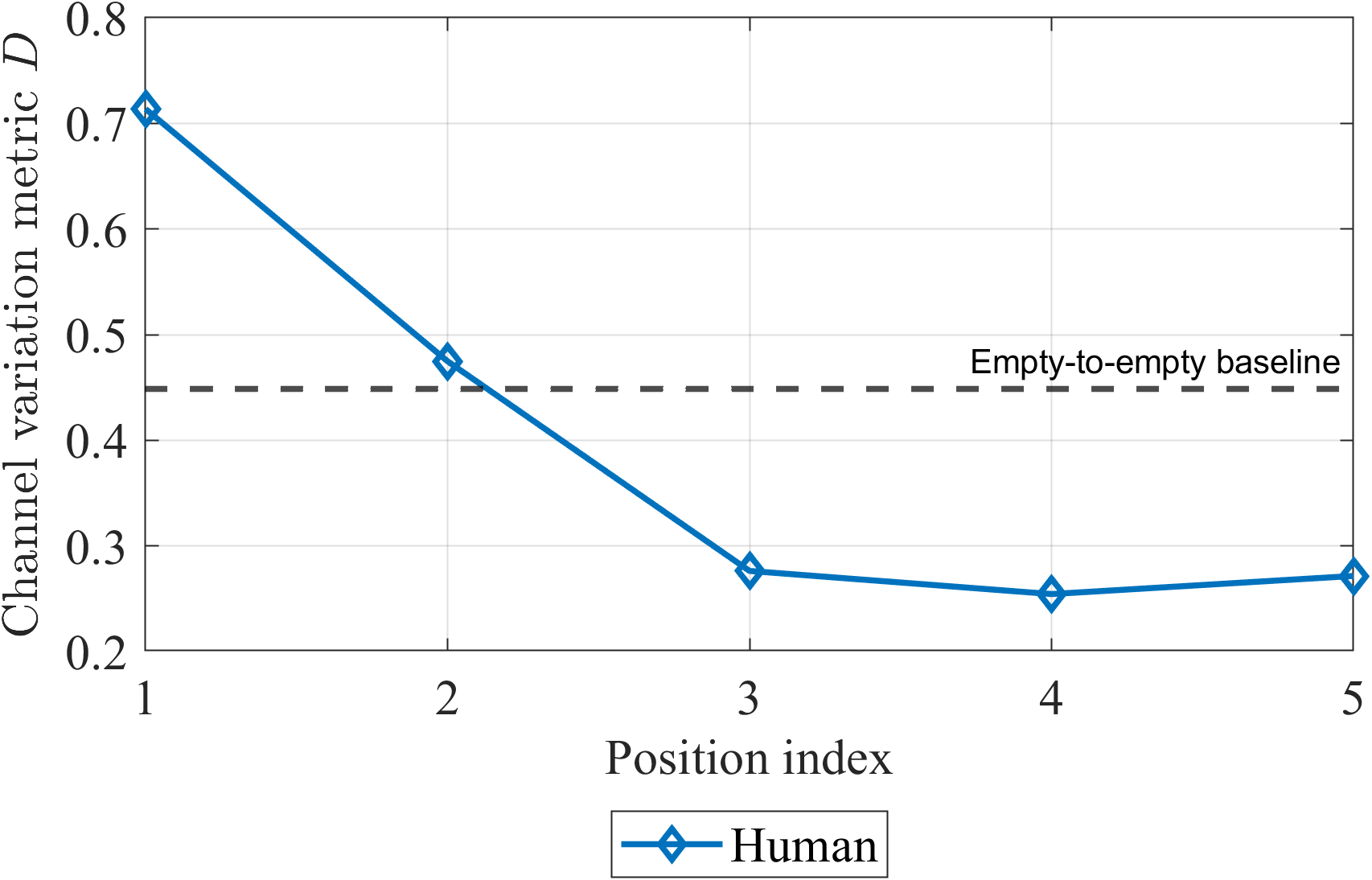}
\caption{Position-wise normalized channel variation metric $D$ for the directive-\gls{tx} configuration. The empty-to-empty baseline $D=0.448$ is shown for reference.}
\label{fig:tx25_metric}
\end{figure}

The detection behavior changes substantially when the transmitter open-waveguide is replaced by the directive antenna, as shown in Fig.~\ref{fig:tx25_metric}. In this case, the human-present normalized channel variation metric reaches approximately $D=0.714$ at P1 and $D=0.475$ at P2, both of which exceed the empty-room baseline of $D=0.448$. For P3--P5, however, the response drops to approximately $D=0.276$, $D=0.254$, and $D=0.272$, respectively, all below the baseline. This demonstrates that the directive transmitter improves detection for favorable target locations, but also makes the sensing coverage more spatially selective. A dense deployment of multiple directive ceiling-mounted transmitters could reduce this coverage limitation by illuminating different regions of the room. However, such a solution would increase hardware complexity and require careful coordination to limit inter-beam or inter-access-point interference \cite{8986750}.

\begin{figure}[!t]
\centering
\includegraphics[width=0.95\columnwidth]{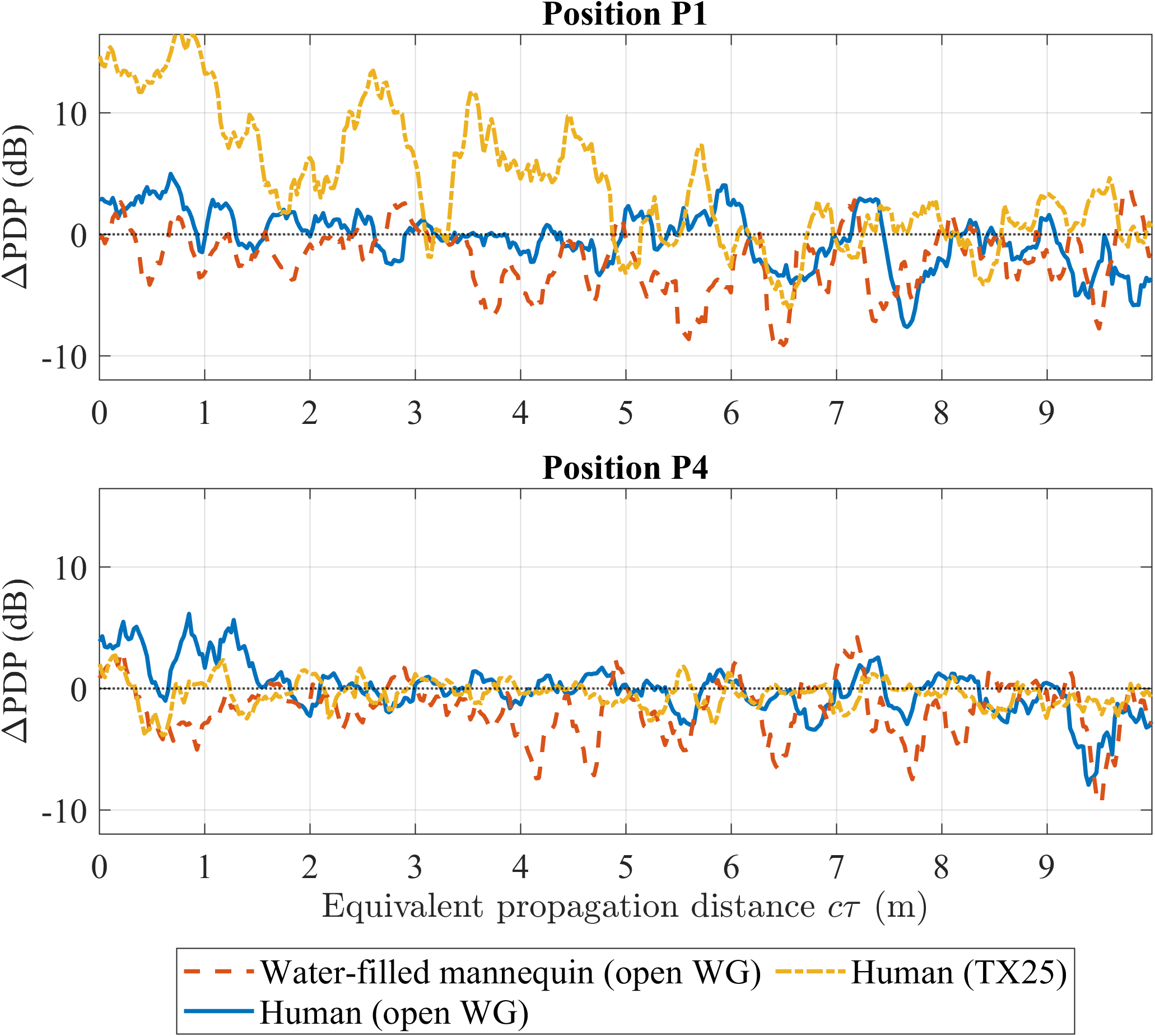}
\caption{Delay-domain \gls{pdp} difference, in dB, relative to the corresponding empty-room reference for two representative positions. Position P1 illustrates a stronger-response case, whereas P4 illustrates a weaker-response case.}
\label{fig:delay_selected}
\end{figure}

The delay-domain behavior underlying these differences in $D$ is shown in Fig.~\ref{fig:delay_selected} for two representative positions. The figure shows the delay-wise difference, in dB, between each target-present \gls{pdp} and its corresponding empty-room reference \gls{pdp}, that is, the target-induced change in the delay-domain response relative to the empty-room case. Position P1 was selected as a stronger-response case because it exhibits the largest human-induced channel variation, whereas P4 was selected as a weaker-response case because it lies among the positions with the smallest observed variation. At P1, all three target-present cases show clear deviations from their empty-room references, with the strongest deviation observed for the directive-\gls{tx} case. At P4, the three curves remain closer to the zero-reference line, indicating weaker target-induced delay-domain changes that are less distinguishable from the baseline channel variation. This behavior agrees well with the position-wise interpretation of $D$ in Figs.~\ref{fig:openwg_metric} and \ref{fig:tx25_metric}, and further shows that the most informative target-induced signatures are localized redistributions of delay-domain energy rather than uniform offsets across the entire \gls{pdp}.

Table~\ref{tab:main_summary_metrics} summarizes the target-present metrics for the human and mannequin measurements. In addition to the normalized channel variation metric $D$ in \eqref{eq:D_metric}, it includes the mean power change $\Delta P_{\mathrm{dB}}$ in \eqref{eq:deltaP_metric}, the mean delay $\bar{\tau}$ in \eqref{eq:mean_delay}, and the \gls{rms} delay spread $\sigma_\tau$ in \eqref{eq:rms_delay}. The column EE $D$ denotes the empty-to-empty baseline for the corresponding antenna configuration.
\begin{table}[!t]
\caption{Summary of target-present metrics. EE denotes the empty-to-empty baseline for the corresponding antenna configuration.}
\centering
\setlength{\tabcolsep}{2.5pt}
\begin{tabular}{p{1.85cm} c c c c c c}
\hline
Scenario & Mean $D$ & Std $D$ & EE $D$ & $\Delta P$ [dB] & $\bar{\tau}$ [ns] & $\sigma_\tau$ [ns] \\
\hline
Human, open WG & 0.299 & 0.061 & 0.455 & -0.164 & 14.653 & 8.623 \\
Mannequin, open WG & 0.497 & 0.015 & 0.455 & -0.847 & 13.941 & 8.958 \\
Human, TX25 & 0.398 & 0.198 & 0.448 & -0.416 & 17.870 & 7.168 \\
\hline
\end{tabular}
\label{tab:main_summary_metrics}
\end{table}The table shows that the metrics provide complementary information. The metric $D$ describes the delay-domain change relative to the empty-room reference, while EE $D$ gives the corresponding empty-room variation level. The power change $\Delta P_{\mathrm{dB}}$ indicates changes in total received energy, whereas $\bar{\tau}$ and $\sigma_\tau$ describe changes in the delay distribution of the received energy. For example, in the open-waveguide configuration, the human case gives $D=0.299$ with a small mean power change of $\Delta P_{\mathrm{dB}}=-0.164$~dB, indicating that the sensing signature is not simply an overall attenuation. The \gls{rms} delay spread is also slightly larger for the human case, $\sigma_\tau=8.623$~ns, and largest for the mannequin case, $\sigma_\tau=8.958$~ns, consistent with a stronger redistribution of energy over delay. For comparison, the corresponding empty-room \gls{rms} delay spreads are $\sigma_\tau=8.533$~ns for the open-waveguide configuration and $\sigma_\tau=6.672$~ns for the directive-\gls{tx} configuration.

Overall, the results show that the delay-domain response contains target-related information, but that its interpretation is not straightforward. In particular, the human and mannequin cases produce different trends relative to the empty-to-empty baseline, showing that both the magnitude and direction of the change in $D$ can be informative. However, these results also show that delay-domain information alone is not sufficient for robust room-wide human detection in the considered ceiling-mounted scenario.

\FloatBarrier

\section{Conclusion}

This paper investigated human presence detection using ceiling-mounted sub-THz channel sounding in a furnished conference room. The measured responses were transformed into calibrated \glspl{pdp} and analyzed using reference-based delay-domain metrics.

The results show that delay-domain measurements contain target-related information, but that robust human detection cannot be based on a single fixed threshold on the normalized channel variation metric $D$. In the open-waveguide configuration, the human-present measurements remained below the empty-to-empty baseline, whereas the water-filled mannequin remained at or above it across all positions. With the directive-\gls{tx} configuration, the human response exceeded the baseline only at favorable positions, especially P1 and P2.

Future work will investigate joint delay, angular, and Doppler-domain sensing to improve robustness, while accounting for the added challenges in measurement complexity and signal processing.

\bibliographystyle{IEEEtran}

\bibliography{ref} % 
\end{document}